\documentclass[preprint,showpacs,preprintnumbers,amsmath,amssymb]{revtex4}
\usepackage{graphicx}
\usepackage{dcolumn}
\usepackage{bm}
\begin{document}
\title{Tantalum Monoarsenide: an Exotic Compensated Semimetal}
\author{Chenglong Zhang$^{\dag 1}$,Zhujun Yuan$^{\dag 1}$, Suyang Xu$^3$, Ziquan Lin$^4$, Bingbing Tong$^1$,M. Zahid Hasan$^3$, Junfeng Wang$^4$, Chi Zhang$^{1,2}$,Shuang Jia$^{1,2}$}
\affiliation{$^1$ICQM, School of Physics, Peking University, Beijing 100871, China\\
$^2$Collaborative Innovation Center of Quantum Matter, Beijing 100871, China\\
$^3$Joseph Henry Laboratory, Department of Physics,Princeton University, Princeton, New Jersey 08544, USA\\
$^4$Wuhan National High Magnetic Field Center, Huazhong University of Science and Technology, Wuhan 430074, China
}

\begin{abstract}
  Compared with the semiconductors such as silicon and gallium arsenide which have been used widely for decades, semimetals have not received much attention in the field of condensed matter physics until very recently. The realization of electronic topological properties has motivated interest of investigations on Dirac semimetals and Weyl semimetals, which are predicted to show unprecedented features beyond the classical electronic theories of metals \cite{Kane05a,Kane05b,BernevigScience,hasanrmp,Wan_Weyl_2011,PhysRevB.85.035103,Hosurweyl2013,Vishwanath_Rev}. In this letter for the first time we report the electric transport properties of a robust Weyl semimetal candidate proposed by recent theoretical calculations \cite{Huang_TaAs_2014,zhangfangTA}, TaAs.
  Our study shows that this bulk material manifests ultrahigh carrier mobility ($\mathrm{5\times10^5 cm^2/V\cdot{s}}$) accompanied by an extremely large, unsaturated linear magnetoresistance ($\mathrm{MR}$), which reaches 5400 at 10 Kelvins in a magnetic field of 9 Teslas and 2.47$\times$10$^4$ at 1.5 Kelvins in a magnetic field of 56 Teslas.
  We also observed strong Shubnikov-de Haas (SdH) oscillations associated with an extremely low quantum limit ($\sim$8 Teslas). Further studies on TaAs, especially in the ultraquantum limit regime, will help to extend the realization of the topological properties of these exotic electrons.

\end{abstract}

\maketitle
A semimetal has a very small overlap between the bottom of the conduction band and the top of the valence band. In the prototype semimetal bismuth, the concentrations of the conduction electrons and holes are extremely low and the Fermi surface occupies a very tiny area of the Brillouin zone. As a result, the quantum limit in bismuth can be reached by the application of a magnetic field as small as 9 Teslas (T)  \cite{behnia2007signatures}. To this date, bismuth still seems to be a unique platform for investigating highly correlated quantum fluid in 3D electron systems \cite{behnia2007signatures,li2008phase}, like the high-mobility semiconductor heterostructures in 2D electron systems. Here we show that a binary compound TaAs, which crystallizes in a very different structure, is another candidate with even more exotic transport properties.

Tantalum monoarsenide crystallizes in a body-centered tetragonal structure with the space group I$4_1md$ \cite{Murray_TaAs_1976},
a non-centrosymmtric structure that is quite rare in transition metal monopnictides in contrast to the ordinary NiAs-type \cite{Johrendt2013}. It can be visualized as an arrangement of regular trigonal prisms of Ta atoms with an arsenic atom at the center. There are four trigonal prisms per unit cell. They are stacked in the way which each prism is rotated and shifted with respected to its nearest neighbour \cite{transposition_TaAs_1963}. The absence of a horizontal mirror plane in the unit cell makes this structure break the inversion symmetry, which is crucial for realizing a time reversal symmetric Weyl semimetal \cite{Huang_TaAs_2014,zhangfangTA}.
Previous resistance measurement on polycrystals showed semimetal- or semiconductor-like behaviors associated with a moderate magnetoresistance $\mathrm{MR=1.1}$ ($\mathrm{MR} = \Delta \rho /\rho _{\mathrm{H}=0}$) at 2 K in 8 T \cite{Sefat_TAs_2012}. These measurements on polycrystals cannot reveal the intrinsic properties of TaAs, because the impurities such as TaAs$_2$ and Ta$_2$O$_5$ on the grain boundaries strongly affect the scattering processes \cite{Sefat_TAs_2012}. Therefore the synthesis of high-quality single crystals is crucial for unveiling the exotic electron behaviors in TaAs.

The temperature dependent resistivity of TaAs at zero magnetic field (Fig. 1A) shows a metallic profile with a residual resistance ratio ($\mathrm{RRR=45}$) for the sample S1. Remarkably, when a low magnetic field (0.3 T) is applied, the resistivity changes to an insulating profile. This behavior is reminiscent of what has been observed in bismuth and graphite, where the low-field effect is attributed to a magnetic-field-induced excitonic insulator transition of Dirac fermions \cite{MITchinese}. We point out that this behavior is distinctive from Cd$_3$As$_2$ \cite{liang2014ultrahigh}, in which the `turn-on' temperatures (defined as the temperatures when $\mathrm{d\rho/dT}$ changes the sign) in different magnetic fields are much lower \cite{liang2014ultrahigh}. Figure 1B shows that the $\mathrm{MR}$ of TaAs shows extremely strong Shubnikov-de Haas (SdH) oscillations below 10 K. The $\mathrm{MR}$ reaches 5400 at 10 K in 9 T, three times larger than that in the recently reported compound WTe$_2$ at 2 K in 9 T \cite{ali_WTe2_2014}. Although the $\mathrm{MR}$ of TaAs is the largest, the $\mathrm{RRR}$ of TaAs (45 for S1) is 1$\sim$2 orders of magnitude less than the values of the recently reported compounds with large $\mathrm{MR}$ \cite{liang2014ultrahigh,Mun_PtSn4_2012,wang2014anisotropic,PdCoO,ali_WTe2_2014}. In Fig. 1C, the large $\mathrm{MR}$ in TaAs follows its linearity ($\mathrm{MR\sim{H^m}}$, H\textgreater0.5 T) ; $\mathrm{m=0.96}$) and shows no indication of saturation at 1.5 K within 56 T (measured on the sample S9), the highest magnetic field we have applied.

The inset of Fig. 1B shows that the $\mathrm{MR}$ of TaAs changes to a parabolic dependence with the strength of the magnetic field and reaches 2.9 at the room temperature. This large $\mathrm{MR}$ at the room temperature is comparable to that of the well-known quantum material silver chalcogenides \cite{xu_large_1997}. The field dependence of the $\mathrm{MR}$ at different temperatures is shown in a logarithmic plot in Fig. 1D. The $\mathrm{MR}$ at low temperatures changes from a parabolic to a linear dependence with a crossover manner characterized by a `turn-on' magnetic field $B^*$. The $B^*$ values change from 0.5 T at 2 K to 6 T at 75 K. Recent theoretical works predict a possible metal-insulator transition and the change of the power law dependence of MR in different strength of magnetic fields in a Weyl semimetal \cite{MIT_Ziegler,Navneeth3Dweylmagneto}. All above indicate that this exotic giant MR in TaAs needs further elaborations.

Figure 1E shows the Kohler's plot at different temperatures (assuming a parabolic dependence in low fields for all tempeartures). According to the Kohler's rule \cite{ziman1972principles}
\begin{equation}\label{1}
 \frac{\Delta\rho_{xx}(T,H)}{\rho_{xx}(T,0)}=F(\frac{H}{\rho_{xx}(T,0)}),
\end{equation}
the $\mathrm{MR}$ measured at different temperatures will fall on the same curve if only one single relaxation time exists \cite{ziman_elePhon_1969} in the scattering process. In Fig. 1E, the $\mathrm{MR}$ violates Kohler's rule over the whole fields, which indicates that different scattering mechanisms must coexist in TaAs. The Hall measurements and the analysis of the SdH oscillations help to understand this complexity.

 The Hall resistivity of TaAs is positive, linearly dependent to the magnetic field up to 9 T above 150 K (Fig. 2A). For temperatures below 100 K,  large, negative field-dependent signals that arise from $n$-type carriers occur at high magnetic fields. Below 20 K, the Hall signals are dominated by strong SdH oscillations,  while the non-oscillatory part shows a large and negative linear-field dependence with a small flat non-linear regime at low fields. We define the `slope-change' field $B^+$ above which the Hall signals become negative linear-field-dependent. As shown in the inset of Fig. 2B, the values of $B^+$ are very close to the values of $B^*$ at different temperatures. This correlation between the Hall resistance and MR indicates that the large linear MR at low temperatures originates from the scattering of an electron band with a low carrier concentration.

 In order to analyze the two types of carriers in TaAs, we calculate the Hall conductivity tensor as $\mathrm{\sigma_{xy}=\rho_{yx}/(\rho_{xx}^2+\rho_{yx}^2)}$, and fit it by adopting a two-carrier model derived from the two-band theory \cite{hall_alloy}
\begin{equation}\label{2}
 \sigma_{xy}=[n_h\mu_h^2\frac{1}{1+(\mu_hH)^2}-n_e\mu_e^2\frac{1}{1+(\mu_eH)^2}]eH,
\end{equation}
Here $n_e$ ($n_h$) and $\mu_e$ ($\mu_h$) denote the carrier concentrations and mobilities for the electrons (holes), respectively. Considering the high power terms \cite{ziman_elePhon_1969}, this model can be used in a full regime of the magnetic fields. Figure 2B gives the fitting at four representative temperatures. As shown in Fig. 2C, the Hall conductivity is mainly from a hole band with negligible electron contribution above 100 K. The $n_h$ decreases and the $\mu_h$ increases with decreasing temperatures. This behavior is consistent with a framework in which the thermal excited $p$-type carriers are scattered by the phonons at high temperatures. Below 100 K, an electron contribution arises associated with a dramatically increasing $\mu_e$ with decreasing temperatures. The $n_e$ and $n_h$ are well-compensated below 75 K while the $\mu_e$ is 5$\sim$10 times higher than the $\mu_h$ at low temperatures. Because of the large $\mu_e$, the values of the $\mu_h$ cannot be estimated accurately in this measurement. The ultrahigh mobility of the electrons ($\mathrm{\mu_e=5\times10^5cm^2/V\cdot{s}}$) at 2 K is comparable with that in Cd$_3$As$_2$ \cite{liang2014ultrahigh} and Bi \cite{moinmetals}. The low carrier concentration makes this binary semimetal unique: as we can see below, the electrons in TaAs can be  driven to these low lying Laudau Levels ($\mathrm{LL}$s), even beyond the quantum limit in a moderate magnetic field.

The measurements of the Hall resistance below 2 K (on the sample S1A) reveal a series of quasi-plateaus corresponding to discrete $\mathrm{LL}$s (Fig. 2D). The first derivative of -$\rho_{yx}$ shows very shape peaks correlated with the maxima of the oscillatory part of $\rho_{xx}$ (the inset of Fig. 2D). The value of $\rho_{yx}$ at the main plateau for n=1 equals 1.9 Ohm-cm, which is 60\% of the Hall resistance quanta $h/e^2$ timing the lattice parameter c (11.656 $\AA$). Exploring the quantum effect in TaAs at the base temperature, in particular in the magnetic field higher than 9 T, is the aim of our study in the next step.

In order to analyze the SdH oscillations of $\rho_{xx}$ in TaAs, we use  the following expression for a 3D system \cite{murakawa2013detection}:
\begin{equation}\label{6}
 \rho_{xx}={\rho_0}[1+A(B,T)cos2\pi(F/B+\gamma)],
\end{equation}
where $\rho_0$ is the non-oscillatory part of the resistivity, $A(B,T)$ is the amplitude of SdH oscillations, $B$ is the magnetic field, and $\gamma$ is the Onsager phase. $\mathrm{F=\frac{\hbar}{2\pi{e}}A_F}$ is the frequency of the oscillations, where A$_F$ is the extremal cross-sectional area of the Fermi surface (FS) associated with the LL index n, $e$ is the electron charge, and $h$ is the Planck's constant($\hbar = h/{2\pi}$). As commonly indexed in a 3D system, the peak and valley positions in $\Delta\rho_{xx} = \rho_{xx}-\rho_0$ are assigned by the integer and half integer indices (because of $\sigma_{xx}\textgreater\sigma_{xy}$), respectively (Fig. 3A). Figure 3B shows the LL indices for four samples, whose frequencies range from 7.42 T to 12.48 T. The interpolation lines of 1/$B$ versus n give the intercepts $\gamma$ near to zero, a non-trivial Berry's phase associated with Dirac fermions. Although the Zeeman splitting of $\mathrm{n=1}$ and 2 peaks is obvious at low temperatures (see Fig. 3F), the LL indices of the sample S1 follow the linearity up to the quantum limit ($\mathrm{n=1}$).

More information about the oscillations of $\rho_{xx}$ for H//$\textbf{c}$ in the sample S1 is obtained by using the following Lifshitz-Kosevich formula for a 3D system \cite{murakawa2013detection}:
\begin{equation}\label{5}
 A(B,T)\propto exp(-2\pi^2k_BT_D/\hbar\omega_c)
 \frac{2\pi^2k_BT/\hbar\omega_c}{sinh(2\pi^2k_BT/\hbar\omega_c)},
\end{equation}
where $k_B$ is  Boltzmann's constant, T$_D$ is the Dingle temperature, and the cyclotron frequency is $\omega_c$ = eB/$m_{cyc}$. Although the FS (A$_F$ = 7.07$\times$10$^{-4}$ $\AA^{-2}$) is comparably small as that in bismuth \cite{moinmetals}, the cyclotron mass $m_{cyc}$ of TaAs is 0.15$m_e$ (Fig. 3C),  much larger than that in Cd$_3$As$_2$ and bismuth ($m_{cyc}$ = 0.044$m_e$ \cite{SYLi} for Cd$_3$As$_2$ and $m_{cyc}$ = 0.008$m_e$ for bismuth along the bisectrix \cite{moinmetals}). Such heavy mass may be correlated with Ta's 5d electron characteristics. The Fermi wave vector $k_F$  is  $(\frac{A_F}{\pi})^{1/2}$ = 0.015 $\AA^{-1}$, and the fermi velocity $v_F$  is $\frac{{\hbar}k_F}{m_{cyc}}$ = 1.16$\times10^5$ $\mathrm{m/s}$. The fermi energy $E_F$ = $m_{cyc}v_F^2$ = 11.48 meV is extremely low, which indicates that the Fermi level is very close to the zero density of states (DOS) points in the momentum space. If we assume an ellipsoid FS (shown in the following part), the electron density $n$ = $k_{Fa}^2k_{Fc}/3\pi^2\approx2.65\times10^{17}$ cm$^{-3}$ ($k_{Fa}$ and $k_{Fc}$ represent the Fermi wave vectors long \textbf{a}-axis and \textbf{c}-axis, respectively) is a factor of three smaller than $n_e$ at 2 K.

In order to better understand the giant $\mathrm{MR}$ in TaAs, we calculated the scattering relaxation time $\tau_Q$ (named as quantum lifetime hereafter) from $\tau_Q$ = $\frac{\hbar}{{2\pi}k_BT_D}$ = $4.73\times10^{-13}$ $\mathrm{s}$ by fitting the Dingle temperature T$_D$ (T$_D$ = 2.6 K) (Fig. 3D). The transport lifetime $\tau_{tr}$ (from standard Bloch-Boltzmann transport) is estimated from the expression $\tau_{tr}$ =  $\frac{\mu_e{\hbar}k_F}{ev_F}$ = $4.51\times10^{-11}$ $\mathrm{s}$, leading to a transport mean free path $l_{tr} = v_F\tau_{tr}\approx5.2$  $\mu{m}$. It is well-known that $\tau_{tr}$ measures backscattering processes that relax the current while $\tau_Q$ is sensitive to all processes that broaden the LLs. The large ratio $R_\tau\equiv\tau_{tr}/\tau_Q$ = 96 in TaAs indicates that the small-angle scatterings are dominant while the backscatterings are strongly protected at low temperatures. The dramatic drop of the $\mu_e$ above 75 K hints that the protection is strongly influenced by the thermal excitation. The giant linear $\mathrm{MR}$ at low temperatures may originates from the removal of the protection by a magnetic field.

Figure 3E shows the angular dependence $\mathrm{MR}$ at 2 K (the angle $\Phi$ is illustrated in the inset of Fig. 3E). Before this measurement, we heated the samples to 200$^\circ$C to solidify the silver paste. As a result, the $\mathrm{MR}$ and the amplitude of SdH oscillations damped. The $\mathrm{MR}$ decreases rapidly when the magnetic field tilted from $\textbf{c}$ to the direction of the current \textbf{a}. In order to map out the shape of the FS, we determine the frequencies F of the SdH oscillations by the analysis of the negative second derivative of resistivity ($-d^2\rho_{xx}/dH^2$) at different angles (Fig. 3F). The inset of Fig. 3E shows that the shape of the FS is ellipsoid-like with its long-axis along the \textbf{c} direction and the largest cross-area is about 40 T.

It has been theoretically predicted that TaAs has multiple gapless Weyl nodes, which are robust enough against any small perturbation of the impurities \cite{Vishwanath_Rev}. In a Weyl system, when the current is parallel to the direction of the magnetic field, the current pumping effect between the Weyl nodes with different chiralities will induce negative longitudinal $\mathrm{MR}$  \cite{abj1983,ABJbisb2013,AjiABJ2012}. The observation of the non-trivial Berry's phase and the giant $\mathrm{MR}$ associated with unexpected large $R_\tau$ in this non-centrosymmetric semimetal TaAs, motivates us to search the signal of the existence of Weyl fermions.

In order to minimize the classical side effects in the longitudinal $\mathrm{MR}$ measurements \cite{mrinmetals,ag2seteprl2005}, we prepared the samples to be long, thin bars (thickness\textless100 $\mu{m}$) with four silver paste contacts fully crossing their wideness. Figure 4A shows the $\mathrm{MR}$ for the sample S3 at 2 K when we stepped the angle $\Phi$ in 2-3$^\circ$ steps through 90$^\circ$. Considering the large positive transverse $\mathrm{MR}$ as a background, we observed a clear minimum $\rho_{xx}$, which we identified as the longitudinal $\mathrm{MR}$ for $\Phi$ = 90$^\circ$. Figure 4B shows that the dip in the longitudinal $\mathrm{MR}$ for H\textless1 T at 2 K changes to a broader feature with increasing temperatures, and finally vanishes above 75 K. A similar profile of the longitudinal $\mathrm{MR}$ in Bi-Sb alloy was explained as the combination of the positive and negative $\mathrm{MR}$ due to the anti-localization effect and chiral effect, respectively \cite{liliang}. Our data fit well with the model, but we believe that more evidences are needed to clarify the Weyl nodes in such complicated system.

Tantalum monoarsenide presents many interesting features of electric transport properties that have never been observed in other semimetals. We have observed the anomalies in the Hall resistance corresponding to several fractional quantum numbers in our first step measurement. The study of the angle-resolved photoemission spectroscopy (ARPES) will help to understand the nature of this exotic semimetal.

\begin{acknowledgments}
We thank Jun Xiong and Fa Wang for valuable discussions. We also thank Yuan Li for using the instruments in his group.
S. Jia is supported by National Basic Research Program of China (Grant
Nos. 2013CB921901 and 2014CB239302).

\end{acknowledgments}


\begin{figure}[p]
  \begin{center}
  \includegraphics[clip, width=1\textwidth]{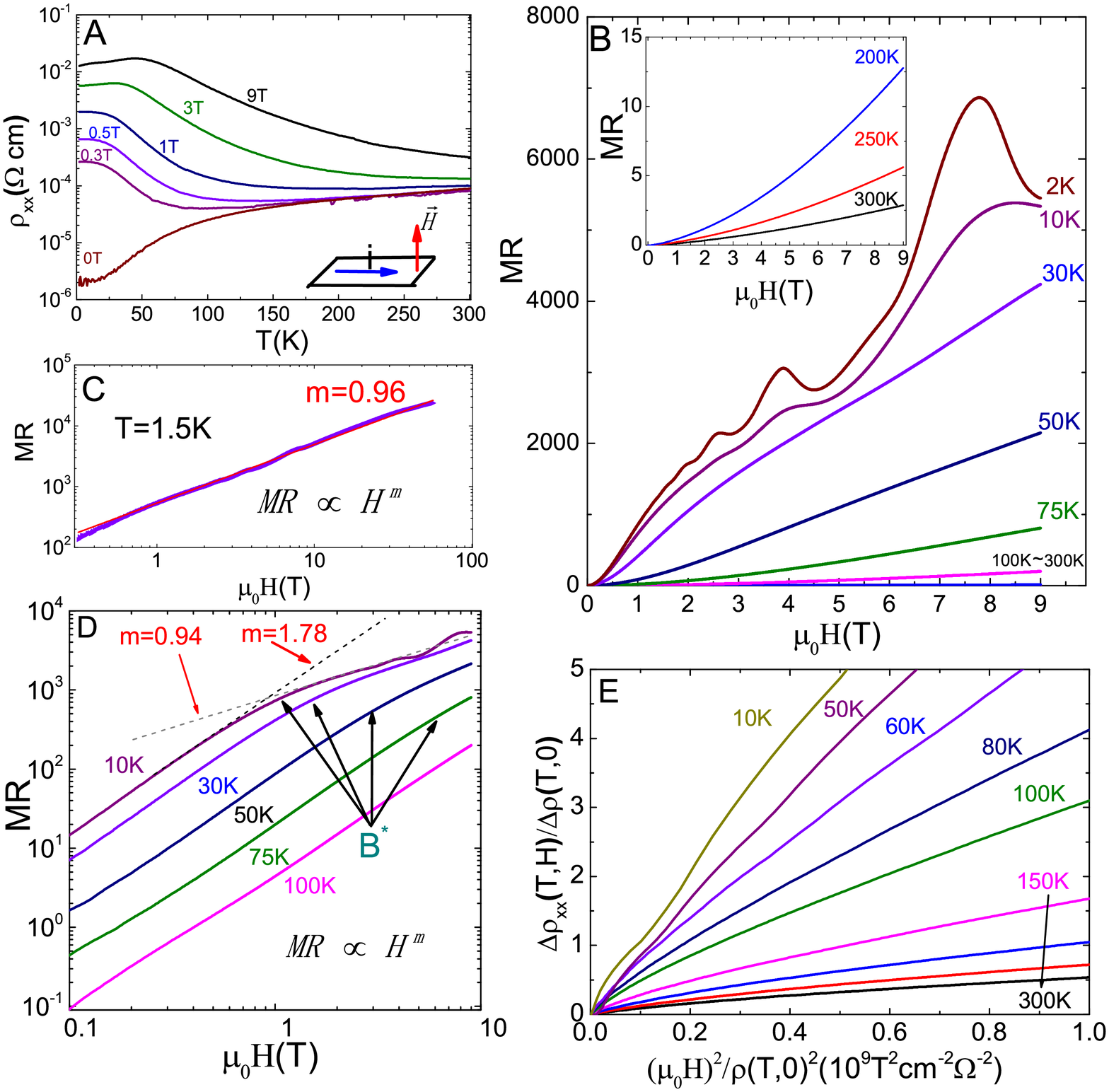}\\
  \caption{\textbf {MR when the magnetic field along the \textbf{c}-axis for TaAs.}  All data are from the sample S1 except that in Panel C. Panel A: the temperature dependent resistivity in different magnetic fields. The experimental setup is shown in a simple sketch as an inset. Panel B: MR at different temperatures. Inset: MR at 200 K, 250 K and 300 K. Panel C: MR of the sample S9 in a pulse magnetic field as high as 56 T. The red line is the fitting result. Panel D: a double logarithmic plot of MR in the regime of 10 K$\sim$100 K. The two solid fitting lines show the different slopes of MR at low and high fields (m = 1.78 in low fields and m = 0.94 in high fields). The `turn-on' fields ($B^*$) are defined as the `slope-change' points on the curves. Panel E: the Kohler's plot of $\Delta\rho_{xx}/\rho(T,0)$ vs. $(B/\rho(T,0))^2$ for MR from 10 K to 300 K.}
  \label{Fig1}
  \end{center}
\end{figure}

\begin{figure}[p]
  \begin{center}
  \includegraphics[clip, width=1\textwidth]{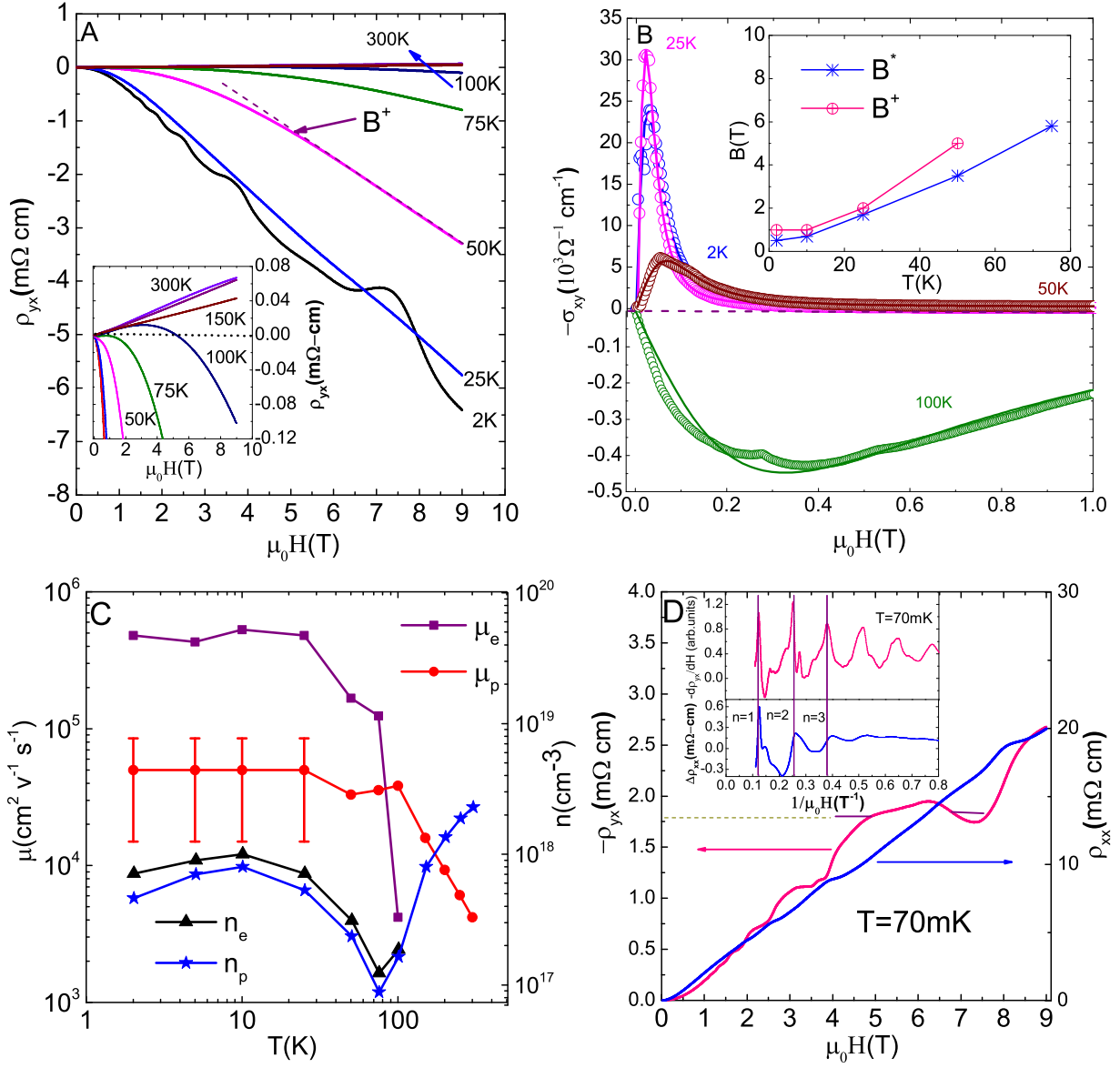}\\
  \caption{\textbf {Hall resistivity and carriers information for TaAs.} All data are from the sample S1 except that in Panel D. Panel A: the Hall resistivity versus magnetic fields from 2 to 300 K. Strong SdH oscillations were observed at 2 K. $B^+$ are defined as the `slope-change' magnetic fields (see more information in the text). Inset: the high temperature Hall resistivity. Panel B: $-\sigma_{xy}$ at four representative temperatures. The solid lines are fitting curves from the two-carrier model. Inset: $B^*$ and $B^+$ at different temperatures. Panel C: the mobilities and carrier concentrations of the electrons and holes. No information of the electrons can be obtained above 100 K in this measurement. Panel D: the -$\rho_{yx}$ and the $\rho_{xx}$ of the sample S1A at 70 mK. Inset: the first derivative of -$\rho_{yx}$ and the oscillatory part of $\rho_{xx}$ at 70 mK.}
  \label{Fig2}
  \end{center}
\end{figure}

\begin{figure}[p]
  \begin{center}
  \includegraphics[clip, width=1\textwidth]{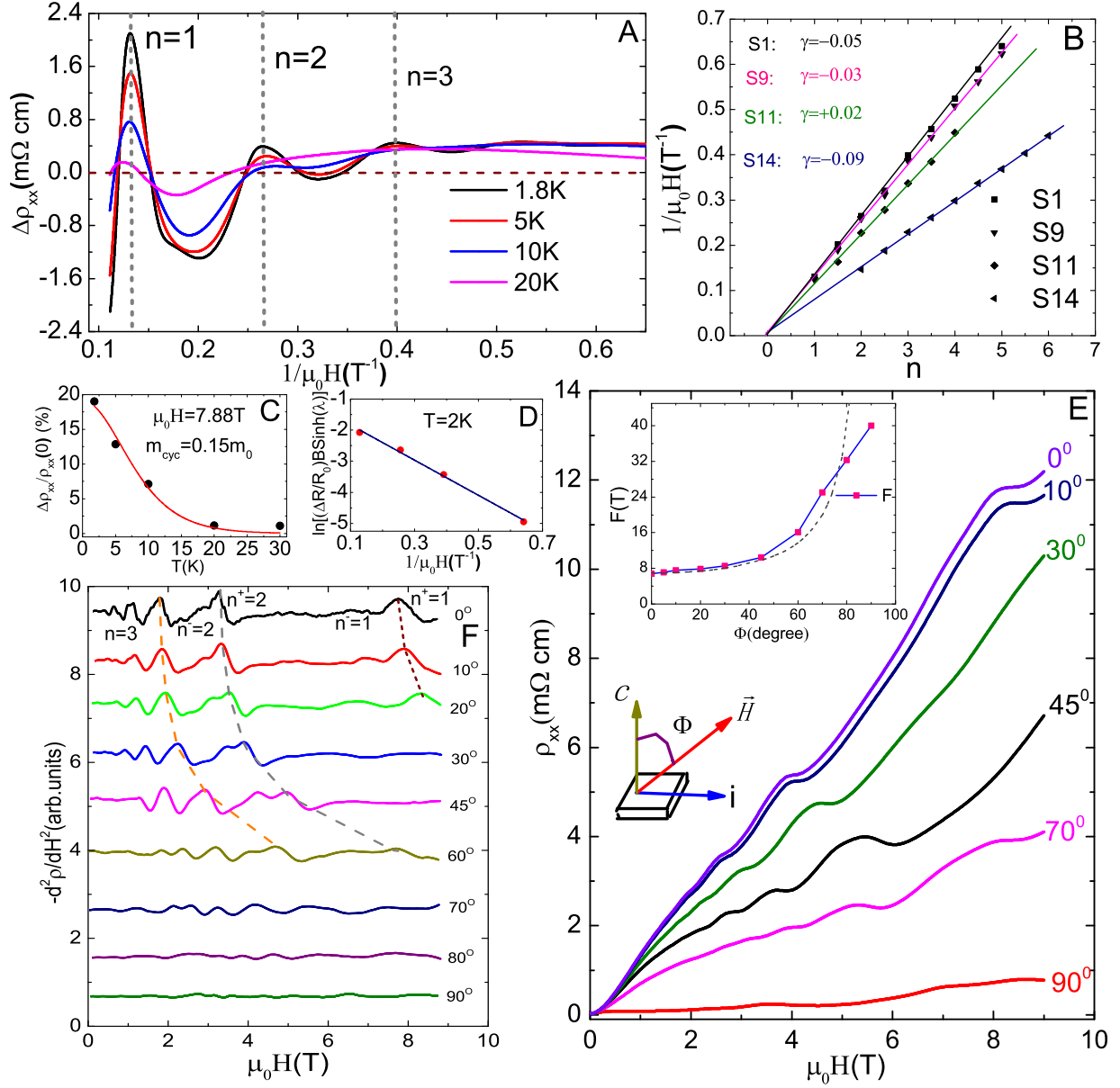}\\
 \caption{\textbf {Analysis of SdH oscillations and mapping for the FS in TaAs.} Panel A: the oscillatory parts of $\rho_{xx}$ at various temperatures for S1. Panel B: the SdH fan diagram for four different samples. All the four intercepts are located around zero. Panel C: the temperature dependent amplitude of the SdH oscillations at 7.88 T for S1. Panel D: the Dingle plot for S1. Panel E: field-dependent resistivity at representative angles at 2 K for S1 after heating (see more details in text). Inset: the frequency F versus $\Phi $. The dashed curve is ($1/cos\Phi$)$\cdot{F_0}$. Panel F: the negative second derivative of resistivity ($-d^2\rho_{xx}/dH^2$) as a function of magnetic fields at 2  K. The Zeeman splitting of n = 1 and 2 is clear.   }
  \label{Fig3}
  \end{center}
\end{figure}

\begin{figure}[p]
  \begin{center}
  \includegraphics[clip, width=1\textwidth]{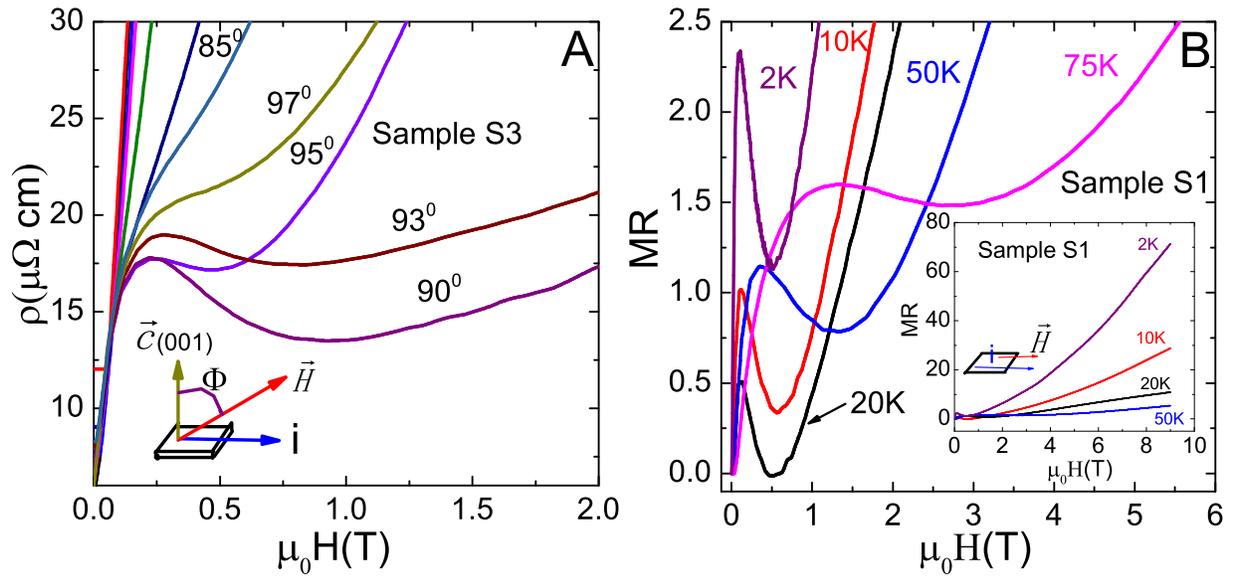}\\
  \caption{\textbf {Longitudinal MR in TaAs.} Panel A: the resistivity of the sample S3 shows a clear dip at low fields at 2 K when $\Phi$ is close to $90^\circ$. Panel B: the longitudinal MR of the sample S1 at different temperatures in low magnetic fields. Inset: the longitudinal MR in high magnetic fields.}
  \label{Fig4}
  \end{center}
\end{figure}

\end{document}